\documentclass[aps,twocolumn,prl,superscriptaddress,amsmath,amssymb]{revtex4-1} 
\usepackage{dcolumn}
\usepackage{bm}
\usepackage{amsmath}
\usepackage{txfonts}
\usepackage[T1]{fontenc}
\usepackage{xspace}
\usepackage{comment}
\usepackage{ulem}
\newcommand{\means}[1]{\langle#1\rangle}
\ifx\pdfoutput\undefined
\usepackage[dvipdfmx]{graphicx}
\usepackage[dvipdfmx]{hyperref}
\usepackage[dvipdfmx]{color}
\else
\usepackage{graphicx}
\usepackage{hyperref}
\usepackage{color}
\fi

\usepackage{natbib}

\usepackage{soul}
\soulregister\cite7
\soulregister\ref7
\soulregister\pageref7

\begin{document}
\title{
Spontaneously orbital-selective superconductivity
in a three-orbital Hubbard model}
\author{Kosuke Ishigaki}
\author{Joji Nasu}
\author{Akihisa Koga}
\affiliation{
  Department of Physics, Tokyo Institute of Technology,
  Meguro, Tokyo 152-8551, Japan
}
\author{Shintaro Hoshino}
\affiliation{
  Department of Physics, Saitama University,
  Saitama 338-8570, Japan
}
\author{Philipp Werner }
\affiliation{
  Department of Physics, University of Fribourg,
  1700 Fribourg, Switzerland
}

 \date{\today}
\begin{abstract}
We study a three-orbital Hubbard model with negative Hund coupling in infinite dimensions,
combining dynamical mean-field theory with continuous time quantum Monte Carlo simulations.
This model, which is relevant for the description of alkali-doped fullerides, has previously been
shown to exhibit a spontaneous orbital selective Mott phase in the vicinity of the superconducting phase.
Calculating the pair potential and double occupancy in each orbital,
we study the competition between different homogeneous ordered states and determine the corresponding
finite temperature phase diagram of the model. We identify two distinct types of spontaneous
orbital-selective Mott states and show that an orbital-selective $s$-wave superconducting state
with one superconducting and two metallic orbitals is
spontaneously realized between the conventional $s$-wave superconducting phase and these
two kinds of spontaneously orbital-selective Mott states.
\end{abstract}
\maketitle


Orbital degrees of freedom and their dynamics are known
to play an essential role in strongly correlated electron systems
as they couple to other degrees of freedom 
of the lattice system.
This can lead to exotic phenomena such as colossal magnetoresistance in
the manganites~\cite{CMR}, 
and unconventional superconductivity in ruthenates~\cite{Maeno} or iron pnictides~\cite{Hosono}.
An interesting phenomenon in this general context is the orbital-selective Mott (OSM) transition~\cite{Anisimov},
which has been discussed in transition metal oxides such as
$\rm Ca_{2-x}Sr_xRuO_4$ \cite{Nakatsuji1,Nakatsuji2} and
$\rm La_{n+1}Ni_nO_{3n+1}$~\cite{Sreedhar,Zhang,Kobayashi}.
The OSM transition results in a distinct electronic character of different orbitals, i.e.,
some orbitals are itinerant while the others are localized.
This physics has been explored in simple two-orbital Hubbard models
with different bandwidths~\cite{KogaOSMT,InabaOSMT,KogaOSMT2,InabaOSMT2,MediciOSMT} or
crystal field splitting~\cite{WernerOSMT},
where the difference of the effective Coulomb interaction or local energy induces the OSM state.

Orbital-selective physics in a model with {\it degenerate} bands is less trivial, 
since it corresponds to a spontaneous breaking of symmetry and an interesting question is whether such
an
OSM transition occurs
simultaneously
with a spontaneous orbital order.
Recently, an exotic state with itinerant and localized orbitals has indeed been observed
in the fullerene-based solids $A_3\mathrm{C_{60}}$ (A=alkali metal) with
triply-degenerate $t_{1u}$ orbitals,
which motivates further theoretical and experimental investigations on
orbital-selective phenomena in such multiorbital systems.
A previous study of a
half-filled three-orbital Hubbard model with antiferromagnetic Hund coupling~\cite{PhysRevLett.118.177002}
revealed the existence of an OSM state with spontaneously broken orbital symmetry (two Mott insulating and one metallic orbital),
and this state has been
referred to as a spontaneously orbital-selective Mott (SOSM) state.
Furthermore, 
it was
demonstrated that this SOSM state is realized
in the vicinity of an $s$-wave superconducting (SC) dome and a Mott insulating phase,
which is consistent with the phase diagram of fullerene-based solids~\cite{Zadik}.
However, these insights were based on susceptibility calculations in the symmetric phase, and
the competition between the SC and SOSM states, as well as the role of orbital fluctuations at low
temperature were not addressed. To clarify these issues it is important to directly examine the symmetry-broken states.

In this Letter, we study the three-orbital Hubbard model
with antiferromagnetic Hund coupling
at half filling,
combining dynamical mean-field theory (DMFT)~\cite{DMFT1,DMFT2,DMFT3}
with continuous-time quantum Monte Carlo (CTQMC) simulations~\cite{CTQMC,CTQMCREV}.
Calculating pair potentials and double occupancies,
we clarify that at low temperatures, an $s$-wave SC state without orbital symmetry breaking is stabilized
rather than the SOSM state.
At higher temperatures, we find a new SOSM state where two orbitals are metallic
while the third is in a paired Mott state.
Most remarkably, we demonstrate that this SOSM phase transforms
into a spontaneous orbital-selective superconducting (SOSSC) phase 
in the vicinity of the SC dome.

We consider the half-filled three-orbital Hubbard model described by the Hamiltonian
\begin{align}\label{3orb}
   \mathcal{H}= &-t\sum_{\means{i,j}\alpha\sigma}c^\dagger_{i\alpha\sigma}c_{j\alpha\sigma}
   +U\sum_{i\alpha}n_{i\alpha\uparrow}n_{i\alpha\downarrow}\nonumber\\
   &+U'\sum_{i\sigma\alpha<\beta}n_{i\alpha\sigma}n_{i\beta\bar{\sigma}}
   +(U'-J)\sum_{i\sigma\alpha<\beta}n_{i\alpha\sigma}n_{i\beta\sigma},
\end{align}
where $c_{i\alpha\sigma}$ is an annihilation operator for an electron
with spin $\sigma(\uparrow,\downarrow)$ and orbital index $\alpha(=1,2,3)$
at the $i$th site and
$n_{i\alpha\sigma}=c^\dagger_{i\alpha\sigma}c_{i\alpha\sigma}$.
$t$ is the transfer integral between nearest neighbor sites,
$U$ $(U')$ is the intraband (interband) Coulomb interaction and
$J$ is the Hund coupling.
We assume the relation $U=U'+2J$
and neglect the exchange part of the Hund coupling and the pair hopping
for simplicity.
The effects of these interactions are discussed later.
In the present calculations, we fix the Hund coupling as $J/U=-1/4$, which is large
compared to ab-initio estimates~\cite{PhysRevB.85.155452} but allows us to reveal the
relevant physics at moderate computational expense.
%
An important point is the negative sign of the (antiferromagnetic) coupling,
which is characteristic of fullerene-based
solids~\cite{Fabrizio,Capone2,Nomura}.
This coupling disfavors singly occupied orbitals since the interorbital Coulomb interaction dominates $(U'>U)$.
At half-filling, the intraorbital Coulomb interaction 
can be regarded as effectively attractive in the weak coupling region.
On the other hand, in the strongly interacting half-filled case,
empty, singly and doubly occupied orbitals are realized 
on a given site
and hence large orbital fluctuations are expected in the system.
This is in stark contrast to the ferromagnetic Hund coupling case with $U>U'$, where
each orbital wants to be singly occupied and orbital fluctuations are suppressed.
Therefore, in our model with antiferromagnetic Hund coupling $(J<0)$,
interesting orbital-selective states may emerge due to
orbital fluctuations.

In the present study, we mainly make use of DMFT.
In this approach, 
 the lattice model is mapped to an effective impurity problem,
where local electron correlations can be taken into account precisely.
The Hubbard model with degenerate orbitals has been extensively discussed
in the framework of DMFT, and interesting phenomena have successfully
been clarified such as
the Mott transition~\cite{Kotliar,Rozenberg,Held,Han,Imai,Koga2band1,Koga2band2,InabaMulti},
orbital-selective Mott transitions~\cite{KogaOSMT,KogaOSMT2,InabaOSMT,InabaOSMT2,MediciOSMT,WernerOSMT,Krylov},
magnetism~\cite{Momoi,Yanatori,Golubeva},
and superconductivity~\cite{PhysRevB.91.085108,Yanatori2}.
In the present study, we focus on the half-filled model and
neglect translational symmetry breaking phases such
as charge density waves, antiferromagnetically or
antiferroorbitally ordered states.
This assumption is justified in a system with
next-nearest-neighbor hopping $t'$ comparable to the nearest-neighbor
hopping $t$~\cite{Chitra}, which should be relevant for fcc-type fullerene-based compounds.

To examine the competition of the SC and SOSM states
at low temperatures,
we calculate the pair potential in the $\alpha$th orbital
$\psi_{\alpha} = |\langle c_{i\alpha\uparrow}c_{i\alpha\downarrow}\rangle |$
as an order parameter of the SC state.
In contrast, the order parameter for the SOSM states is not obvious
since 
no difference in the average orbital occupations appears
\cite{PhysRevLett.118.177002}.
Here, we calculate the double occupancy for orbital $\alpha$,
$
d_\alpha=\langle n_{i\alpha\uparrow}n_{i\alpha\downarrow}\rangle,
$
and characterize the SOSM state by the appearance of orbital-dependent double occupancies.
In the following, we set the unit of energy to the half-bandwidth $D$. 

\begin{figure}[t]
 \begin{center}
 \includegraphics[width=7cm]{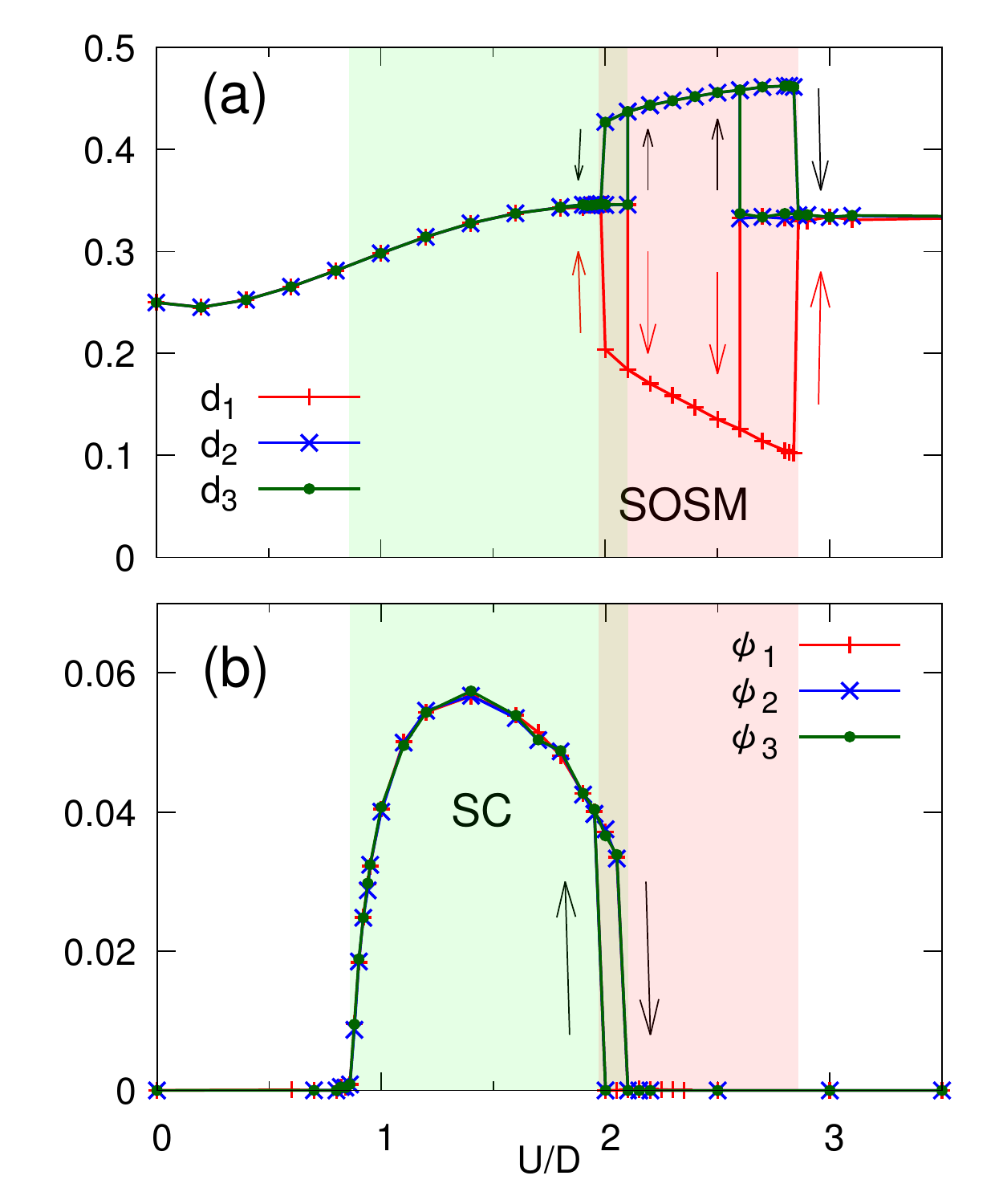}
 \caption{(a) Double occupancy and (b) pair potential as a function of
 the interaction strength $U$ in the three-orbital Hubbard model at
 $T/D=0.01$.
 The arrows indicate the existence of a hysteresis in these quantities.
 }
 \label{SOSMvsSC}
 \end{center}
 \end{figure}

Figure~\ref{SOSMvsSC} plots the pair potential and double occupancy for each orbital
at $T/D=0.01$.
In the noninteracting case ($U=0$),
a metallic state is realized with $d_\alpha=1/4$ and $\psi_\alpha=0$ for all orbitals.
Turning on the interaction $U$ slightly
increases the double occupancy since the onsite interaction in this half-filled system is
effectively attractive due to the antiferromagnetic Hund coupling.
This interaction is expected to enhance pair correlations, and indeed a
second-order phase transition occurs to
the $s$-wave SC state with finite $\psi_1=\psi_2=\psi_3$
around $U/D\sim 0.87$, as shown in Fig.~\ref{SOSMvsSC}(b).

A further increase of the Coulomb interaction leads to a maximum
in the pair potentials near $U/D\approx 1.4$.
Around $U/D\sim 2.0$,
the physical quantities jump and a
first-order phase transition occurs,
as shown in Figs.~\ref{SOSMvsSC}(a) and~\ref{SOSMvsSC}(b).
In the state with $2.0\lesssim U/D \lesssim 2.8$,
the pair potentials vanish, while
the double occupancies take two distinct values
$d_1 < d_2=d_3$.
This means that a metallic state is realized in orbital 1.
In contrast, a paired Mott state with $d\sim 0.5$ is realized in orbitals 2 and 3,
which are dominated by empty and doubly occupied configurations.
From these observations, we can conclude that
an SOSM state is indeed realized in this region~\cite{PhysRevLett.118.177002}.
A similar orbital symmetry breaking has also been identified in the two-dimensional system \cite{Misawa}.
Beyond $U/D\sim 2.8$,
the orbital-selective features disappear, and
the double occupancies exhibit an orbital-independent value.
In the strong coupling region, three electrons are localized at each site
in a configuration with empty, singly, and doubly occupied orbitals,
and a Mott state is realized with $d_\alpha\sim 1/3$.

\begin{figure}[t]
 \begin{center}
 \includegraphics[width=8cm]{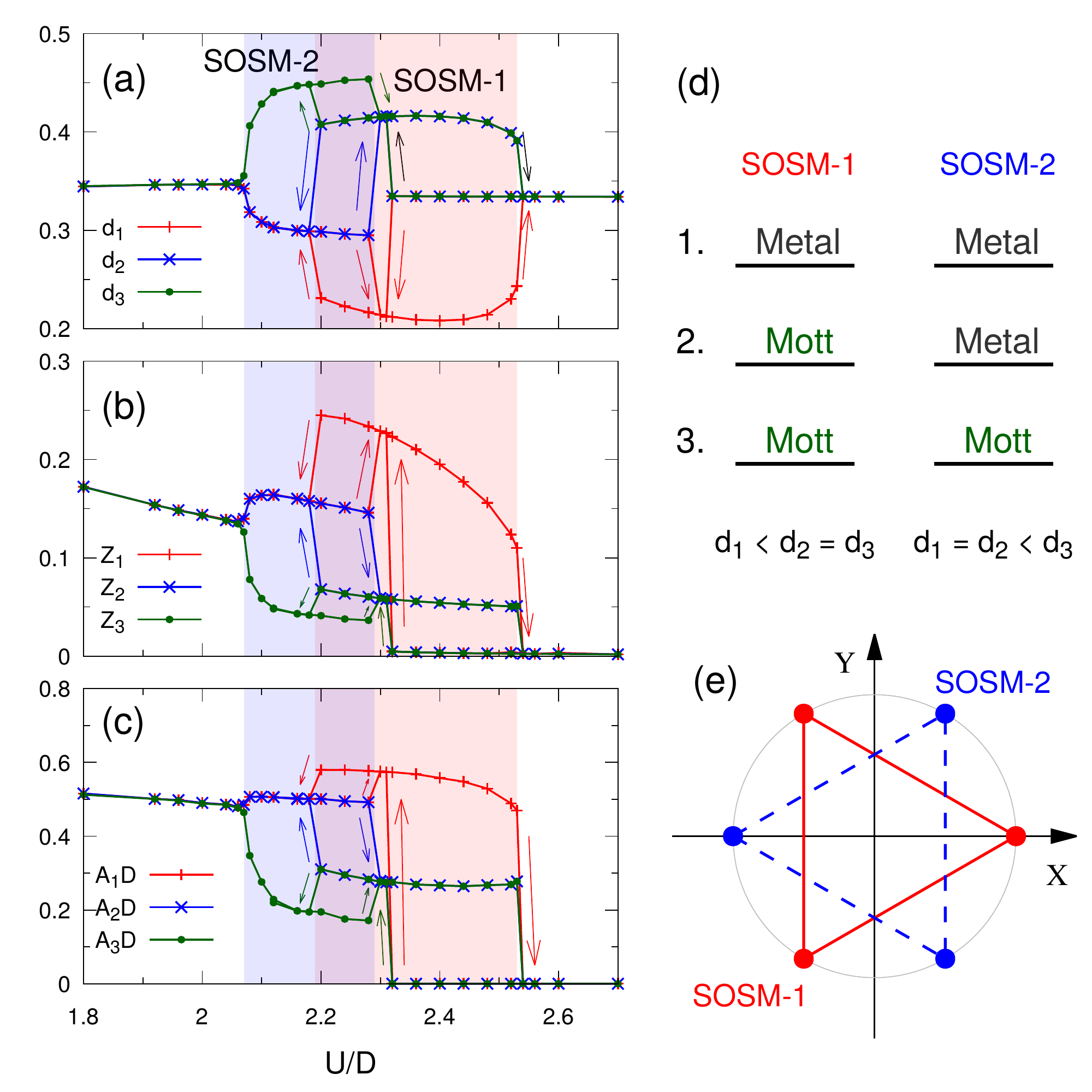}
 \caption{(a) Double occupancy $d_\alpha$, (b) renormalization factor $Z_\alpha$, and
   (c) the quantity $A_\alpha$ as a function of $U/D$ when $T/D=0.02$.
 The arrows indicate the existence of a hysteresis in the quantities.
(d) Schematic pictures for the SOSM-1 and SOSM-2 states.
(e)  Classification of orbital orders in the three-orbital Hubbard model.
The points connected by lines indicate the equivalent solutions.
 }
 \label{SOSM2}
 \end{center}
 \end{figure}
These results confirm that the SOSM state
with one orbital itinerant and two orbitals localized competes with
the $s$-wave SC and Mott states at low temperatures.
On the other hand,
it is naively expected that
another SOSM state with two orbitals itinerant and one orbital paired
may also exist in the present system although such a state has not been
previously discussed~\cite{PhysRevLett.118.177002}.
To clarify this, we examine the low temperature properties
systematically,
calculating the orbital-dependent quantities
$Z_\alpha=[1-{\rm Im} \Sigma_\alpha(i\omega_0)/\omega_0]^{-1}$ and
$A_\alpha=-G_\alpha(1/2T)/\pi T$
as estimates of the renormalization factor and the density of states at Fermi level
at finite temperatures~\cite{DOST},
where $\Sigma_\alpha$ is the self-energy for the $\alpha$th orbital
and $\omega_0=\pi T$.
The results at the temperature $T/D=0.02$ are
shown in Figs.~\ref{SOSM2} (a)-(c).
In this parameter region, no pair potentials appear.
It is found that
two kinds of SOSM states are realized
between the metallic and Mott states.
These can be classified by the number of itinerant unpaired orbitals;
the SOSM-$n$ state is associated with $n$ $(=1,2)$ metallic orbital(s),
which are schematically shown in Fig.~\ref{SOSM2}(d).
When $2.2 \lesssim U/D \lesssim 2.5$,
$d_1<d_2=d_3$ and the SOSM-1 state is realized with a metallic orbital 1.
On the other hand, in the region $2.1\lesssim U/D \lesssim 2.3$,
$d_1=d_2<d_3$ and the SOSM-2 state is realized with metallic orbitals 1 and 2.
The phase transitions between metallic, SOSM-2, SOSM-1,
and Mott states should be of first order although no hysteresis is visible
around $U/D\sim 2.1$.

To clarify the nature of the phase transition,
we employ a Landau theory,
where the symmetry of the system is taken into account correctly.
As discussed above,
the orbital-dependent double occupancies
$d_\alpha$ 
are appropriate to characterize the SOSM states~\cite{PhysRevLett.118.177002}.
Since the Hamiltonian (\ref{3orb}) is invariant under permutations
of the three orbitals,
the free energy $F$ should be expanded as
\begin{align}
  F &= F_0 + a(X^2+Y^2) + bX(X^2- 3Y^2)  + c(X^2+Y^2)^2,
\end{align}
where $X= \left(d_1+d_2 - 2d_3\right)/\sqrt{3}, Y= d_1 - d_2,$
with constants $F_0$, $a$, $b$, and $c(>0)$.
$X$ and $Y$ correspond to the order parameters
characteristic of the SOSM states and
their forms derive from the $3\times 3$ Gell-Mann matrices
$\lambda_8$ and $\lambda_3$, respectively.
The orbital permutation is then represented by
the $C_{3V}$ symmetry in the $X$-$Y$ plane.
This yields the third-order term in the free energy
and the phase transition to the SOSM states is of first order
(related to the Lifshitz condition).

The nontrivial solutions can be classified into two classes.
The solution with $(X,Y)=(-\tfrac{1}{2}R,-\tfrac{\sqrt 3}{2}R)$ ($R>0$ is a radius),
which is equivalent to
$(R,0)$ and $(-\tfrac{1}{2}R,\tfrac{\sqrt 3}{2}R)$ under the $C_{3V}$ symmetry,
corresponds to the SOSM-1 state with $d_1<d_2=d_3$.
The other class is the SOSM-2 solution with $(X,Y)=(-R,0)$,
where $d_1=d_2<d_3$.
These solutions in the $X$-$Y$ plane are schematically
shown in Fig.~\ref{SOSM2}(e).
Namely, the SOSM-1 (SOSM-2) state is stabilized
in the negative (positive) $b$ case.
Therefore, in the system at $T/D=0.02$,
the sign of $b$ changes around $U/D\sim 2.25$.

 \begin{figure}[t]
 \begin{center}
 \includegraphics[width=7cm]{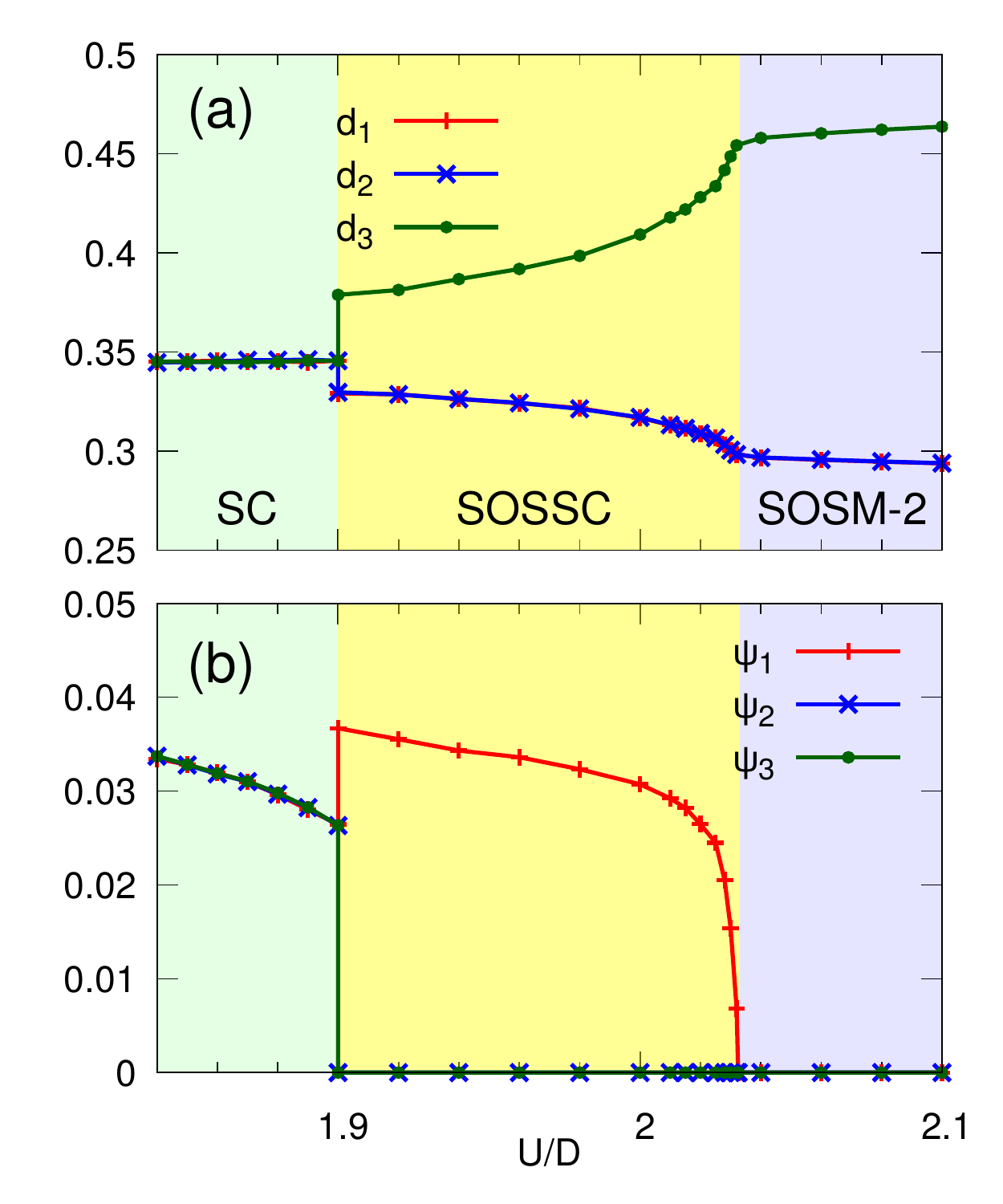}
 \caption{
 (a) Pair potential and (b) double occupancy at $T/D=0.013$.
 }
 \label{SOSSC}
 \end{center}
 \end{figure}
At lower temperatures, an interesting orbital-selective state appears.
Figure~\ref{SOSSC} shows the double occupancy and pair potential
for each orbital when $T/D=0.013$.
When $U/D\lesssim 1.90$, the $s$-wave SC state is realized
with $\psi_1=\psi_2=\psi_3\neq 0$.
In the region with $U/D\gtrsim 2.03$,
the SOSM-2 state is realized with
$d_1=d_2<d_3$ and $\psi_1=\psi_2=\psi_3=0$.
Between these two states $(1.90\lesssim U/D \lesssim 2.03)$,
we find in Fig.~\ref{SOSSC}
that the pair potentials as well as the double occupancies
take two distinct values.
In particular, one of the three orbitals has a finite pair potential
while it vanishes for the other two.
This implies the realization of a spontaneously orbital-selective
superconducting (SOSSC) state.

Now, let us consider the nature of the SOSSC state.
As shown in Fig.~\ref{SOSSC}, the phase transition at $U/D\sim 1.90$ is of first order
whereas that at $U/D\sim 2.03$ appears to be continuous.
This suggests that the SOSSC state is closely related to the higher $U$ state, i.e., the SOSM-2 state.
In this state, the orbital 3 is in a paired Mott state with large $d_3$ and the others are metallic.
Decreasing $U$ from the SOSM-2 phase, $\psi_3$ becomes finite at $U/D\sim 2.03$
with an accompanying rapid decrease of the double occupancy $d_3$.
This indicates that the orbital 3 plays an essential role in the phase transition
to the SOSSC state.
The behavior observed in orbital 3 is similar to the low temperature properties
of the single-band attractive Hubbard model at half filling,
where a second-order phase transition occurs between the SC and
paired Mott states~\cite{KogaAtt1,KogaAtt2}.
This is consistent with the present result that 
 the phase transition at $U/D\sim 2.03$ is of second order.

The orbital-selective superconducting instability originates from
the existence of paired Mott orbitals in the SOSM state.
Therefore, a different type of SOSSC state may exist adjacent to the SOSM-1 state,
whose essential feature should be described in terms of the two-band Hubbard model.
The SC state in the latter model is realized
in a narrow parameter space~\cite{PhysRevB.91.085108},
which suggests that the potential SOSSC state related to the SOSM-1 phase is less stable and
is difficult to realize in the present parameter regime.

\begin{figure}[htb]
 \begin{center}
 \includegraphics[width=8cm]{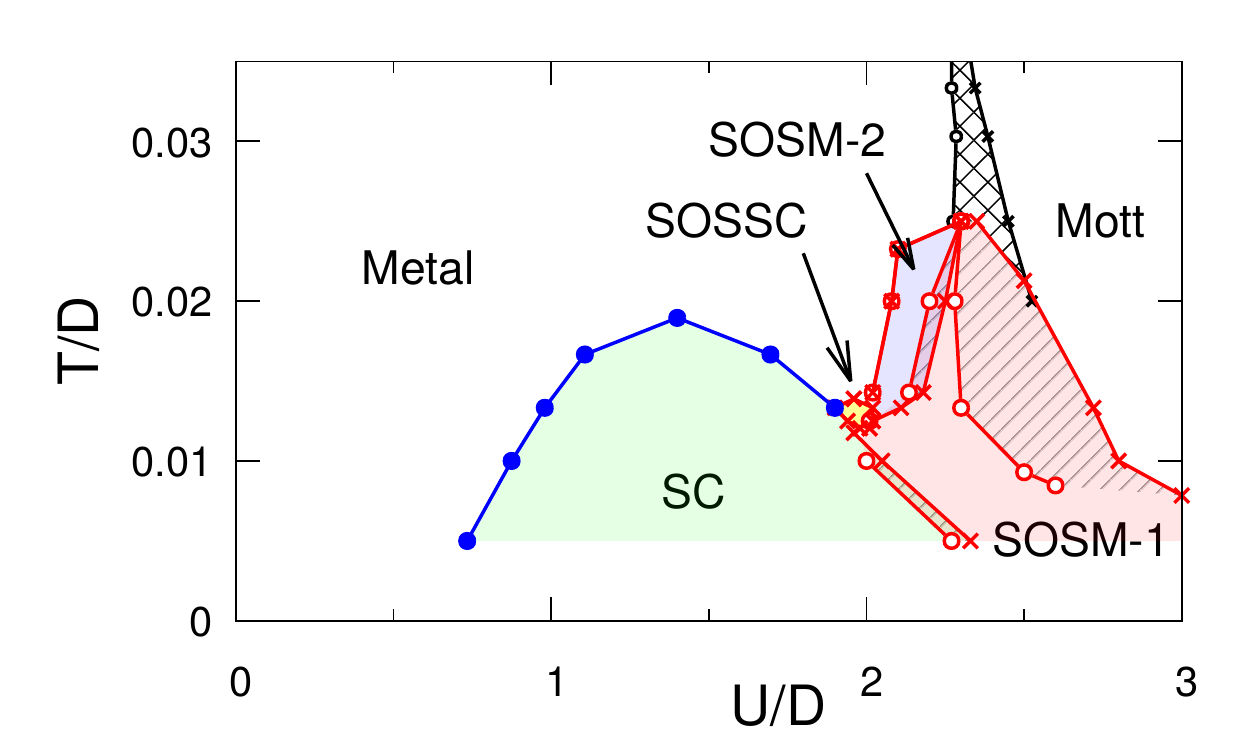}
 \caption{
Phase diagram of the three-orbital Hubbard model.
Solid circles represent second-order phase transition points.
Open circles (crosses) represent the transition points,
where the strong (weak) coupling state disappears.
Shaded areas bounded by these points indicate
the regions with two competing solutions.
 }
 \label{frustPD}
 \end{center}
 \end{figure}
By performing similar calculations for different temperatures,
we obtain the phase diagram shown in Fig.~\ref{frustPD},
which clarifies the competition between the different ordered phases.
For example, this phase diagram indicates that the first-order phase boundary between
 the $s$-wave SC and SOSM-1 states 
shifts to larger $U$ 
with decreasing temperature.
In the stronger coupling region and at low temperatures, the SOSM-1 state is realized
instead of the Mott state adjacent to the SC state.
This means that, due to the breaking of orbital symmetry,
the SOSM-1 state is more stable than the Mott state.

The dominant electronic configurations for the SOSM-1 and SOSM-2 states are similar, but
the SOSM-2 state is stabilized only at higher temperatures.
This can be explained as follows.
The SOSM states possess both itinerant and paired-Mott orbitals.
The entropies for itinerant and paired Mott orbitals should be given by
$S\sim \gamma T$ and $S\sim \ln 2$, respectively,
where $\gamma$ is the specific heat coefficient proportional to
the effective mass.
Therefore, at high temperatures,
the metallic orbitals tend to have a large entropy.
For this reason, the SOSM-2 state with two metallic orbitals is realizable
only at intermediate temperatures.
For similar reasons, the SOSSC state with one orbital superconducting and
the others metallic is less stable than the 
 SC state with all orbitals superconducting at zero temperature.
Therefore, the SOSSC state is stabilized only at finite temperatures.

In the present DMFT calculations,
we did not consider the spin exchange part of the Hund coupling and pair hopping term.
For the system with antiferromagnetic Hund coupling, the low-spin state is favored and the spin-flip is irrelevant.
On the other hand, it has been clarified in Ref.~\onlinecite{PhysRevLett.118.177002} that the pair hopping is relevant
to stabilize the SOSM-1 and SC states. 
To reveal the effect on the SOSM-2 state, we adopt
a phenomenological theory for simplicity
(details are given in the Supplemental Material).
In the case without pair hopping, the SOSM-2 state appears between the high-temperature metallic and low-temperature SOSM-1 states.
This is consistent with the result obtained by the DMFT calculations, which supports the validity of
our phenomenological theory for the present system.
Applying this theory to the system with pair hopping yields the prediction that the SOSM-2 state also exists in the intermediate temperature region.
This
result suggests that the SOSM-2 state survives even in the presence of pair hopping.

In summary,
we have studied the three-orbital Hubbard model in infinite dimensions,
combining DMFT with the CTQMC method.
Calculating the pair potential and double occupancy in each orbital,
we have determined the finite temperature phase diagram of the model.
We have clarified that an orbital-selective $s$-wave superconducting state
with one orbital superconducting and the others metallic
is spontaneously realized, in addition to
the conventional $s$-wave superconducting state and
two kinds of spontaneously orbital-selective Mott states.
Finally, we briefly discuss the relevance to real materials.
The Jahn-Teller metal in the fullerene-based solids $A_3\mathrm{C_{60}}$ is a promising candidate for the SOSM-1 state ~\cite{PhysRevLett.118.177002}.
Our results suggest that the higher-temperature part of the Jahn-Teller metal is in fact an SOSM-2 state.
If one can experimentally distinguish the SOSM-2 state from the low-temperature SOSM-1 state, for example by a difference in the electrical conductivity, the new type of spontaneous orbital selective superconducting state 
should exist in the vicinity 
of this transition line and the SC phase
in $A_3\mathrm{C_{60}}$.

\begin{acknowledgments}
Parts of the numerical calculations were performed
in the supercomputing systems in ISSP, the University of Tokyo.
This work was supported by Grant-in-Aid for Scientific Research from
JSPS, KAKENHI Grant Nos. JP18K04678, JP17K05536 (A.K.),
JP16K17747, JP16H02206, JP18H04223 (J.N.),
JP16H04021 (S.H.)
and the European Research Council through ERC Consolidator Grant 724103 (P. W.).
\end{acknowledgments}

\bibliography{./refs}

\end{document}


\title{
Supplementary material for\\
``Spontaneously orbital-selective superconductivity
in a three-orbital Hubbard model''}
\author{Kosuke Ishigaki}
\author{Joji Nasu}
\author{Akihisa Koga}
\affiliation{
  Department of Physics, Tokyo Institute of Technology,
  Meguro, Tokyo 152-8551, Japan
}
\author{Shintaro Hoshino}
\affiliation{
  Department of Physics, Saitama University,
  Saitama 338-8570, Japan
}
\author{Philipp Werner }
\affiliation{
  Department of Physics, University of Fribourg,
  1700 Fribourg, Switzerland
}

 \date{\today}
\maketitle

In this material we develop a phenomenological theory that accounts for the thermodynamics of SOSM states.
We introduce the labels 0, 1 and 2 to indicate orbital-symmetric metal state (three metallic orbitals), SOSM-1 (one metallic orbital / two paired orbitals) and SOSM-2 (two metallic orbitals and one paired orbital), respectively.
A metallic orbital results in a free energy gain from the kinetic energy ($K$) and the $T$-linear entropy ($S\propto T$).
For a paired orbital, on the other hand, there is a free energy gain from the effective attraction ($V$), and also from the entropy $S=\ln 2$ associated with the degrees of freedom of locating the pairs in the orbitals.
The effective free energies can thus be expressed as 
\begin{align}
\begin{matrix}
F_0 & = & -3K - T(3\gm T)
\\
F_1 & = & -K-2V - T(\ln 2 + \gm T)
\\
F_2 & = & -2K-V - T(\ln 2 + 2\gm T)
\end{matrix}
\ \ \ ,
\end{align}
where $K>0$, $V>0$, $\gm>0$ are the (renormalized) kinetic energy, effective attraction, and specific heat coefficient ($\gm \sim 1/ K$), respectively. 
Our purpose here is to describe the thermodynamic stability of these states, while the kinetic energy of the pairs, which is necessary for e.g. superconductivity, is neglected.
We focus on the low-energy region and neglect the $T$ dependence of $K,V,\gm$.

We are interested in the stability of the SOSM-2 state on which the SOSSC is founded.
The inequalities $F_0 > F_2$ and $F_1 > F_2$ are
the necessary and sufficient condition for the realization of SOSM-2 as the most stable state.
This leads to the relation
\begin{align}
V>K - \frac{(\ln 2)^2}{4\gm}
,
\end{align}
which determines the lower bound of the attractive interaction $V$.
The typical temperature dependences of the free energy are shown in the upper panels of Fig. 1.
When the effective attraction is sufficiently large, the SOSM-2 state becomes the most stable one in the intermediate  temperature range, which is qualitatively consistent with the results shown in the main text.
This demonstrates that the above simple theory can capture the thermodynamics of the SOSM states realized in the DMFT study.

Since the above phenomenology works well for the effective description of the DMFT results, we now apply it to the system with pair hopping.
The free energies in this case are
\begin{align}
\begin{matrix}
F_0 & = & -3K - T(3\gm T)
\\
F_1 & = & -K-2V -J - T(\gm T)
\\
F_2 & = & -2K-V - T(\ln 2 + 2\gm T)
\end{matrix}
\ \ \ ,
\end{align}
where $J>0$ is the (effective) pair hopping.
The pair hopping modifies only $F_1$, since the two paired orbitals are quantum-mechanically mixed, which results in an  energy gain of $J$.
The condition for the situation where SOSM-2 is most stable in the presence of pair hopping becomes 
\begin{align}
&V > K - \frac{(\ln 2)^2}{4\gm}
,\label{eq:1}\\
&J <\frac{\ln 2}{\gm}\left(
\ln 2 + \sqrt{(\ln 2)^2 + 4\gm(V-K)}
\right)
.\label{eq:2}
\end{align}
Thus, there is an upper bound for the magnitude of $J$, in addition to a lower bound for $V$.
The typical temperature dependences of the free energies are shown in the lower panels of Fig. 1, and the necessary conditions described by Eqs.~\eqref{eq:1} and \eqref{eq:2} are indicated in the right-most panel.
Although it is not trivial to determine whether or not the realistic values are located inside of this region,
 there is a chance that the SOSM-2 state is realized if the effective attraction $V$ is strong enough, which pushes the upper bound of $J$ to higher values.


\begin{figure*}[t]
\begin{center}
\includegraphics[width=180mm]{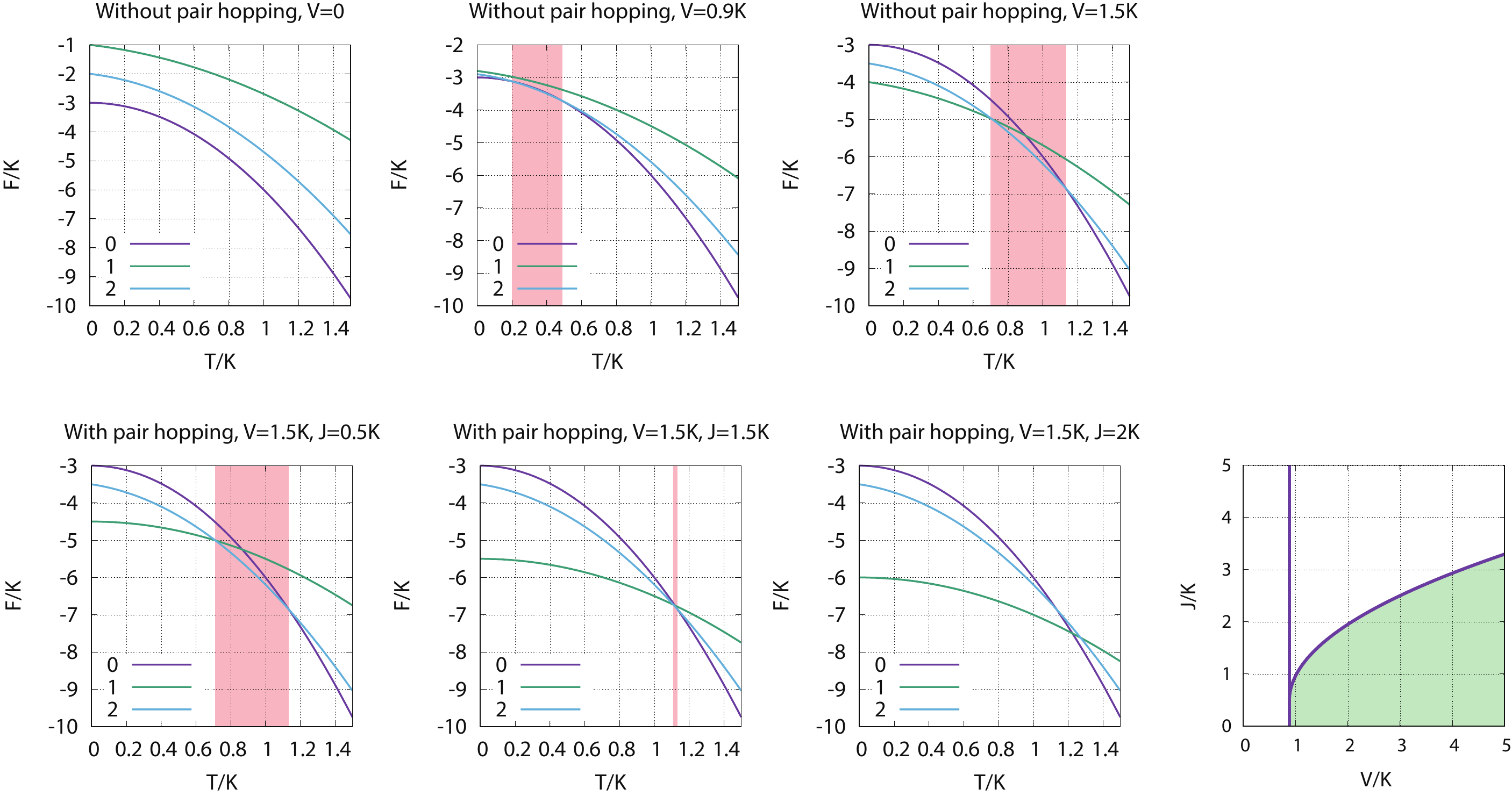}
\caption{
Temperature dependences of the free energies without (top panels) and with (bottom panels) pair hopping.
We have used $\gm = \al/K$ with $\al=1$.
The intervals highlighted in red indicate the temperature range where the SOSM-2 is the most stable.
The right-most panel in the bottom row shows the stability region of the SOSM-2 state determined by Eqs.~\eqref{eq:1} and \eqref{eq:2}.
}
\label{fig:1}
\end{center}
\end{figure*}